\documentstyle[editedvolume,graphics,psfig]{crckapb}

\begin{opening}
\title{Time-scale for accretion of matter}

\author{F. COMBES}
\institute{ DEMIRM, Observatoire de Paris\\ 61 Av. de l'Observatoire,
              F--75014 Paris, France}

\end{opening}

\runningtitle{Time-scale for accretion of matter}

\begin{document}
\section{Abstract}

Mass accretion is the key factor for evolution of galaxies. It can
occur through secular evolution, when gas in the outer parts is driven
inwards by dynamical instabilities, such as spirals or bars. This secular
evolution proceeds very slowly when spontaneous, and can be accelerated
when triggered by companions. Accretion can also occur directly
through merging of small companions, or more violent interaction and
coalescence. We  discuss the relative importance of both processes,
their time-scale and frequency along a Hubble time. Signatures of both
processes can be found in the Milky Way. It is however likely
that our Galaxy had already 
gathered the bulk of its mass about 8-10 Gyr ago, as is expected
in hierarchical galaxy formation scenarios.

\section{Introduction}

 There are two essential processus for a galaxy to accrete mass:
either it accretes gas regularly, through internal dynamics,
producing radial flows, from gas in the outer parts (the reservoir 
could be in the Local Group); or the accretion occurs in more violent
events, galaxy interactions or mergers. The first process, that 
will be called secular accretion, can also be triggered and 
enhanced by the passage of companions, so that the two processus are
in fact inter-related.
Let us try to estimate the corresponding time-scales involved.

\section{Secular evolution}

 The galaxy can be considered as a giant accretion disk: to minimise
its energy, it has the tendency to concentrate, and accrete gas from
the outer parts towards the center. But the angular momentum is
a barrier: only through tangential forces, creating torques,
can the angular momentum be exchanged and transfered outwards
in order that the mass flows inwards. One can apply an analog
of the theory of viscous disks (Lin \& Pringle 1989).
In that frame, viscous torques are the way to re-distribute
angular momentum. If the time-scale for this re-distribution,
$\tau_{vis}$, is of the same order of magnitude as the
time-scale to form stars $\tau_*$, then an exponential 
stellar disk is created, and this generates also an exponential
distribution of metallicity (Tsujimoto et al 1995).

\subsection{Gravity Torques}

What is the nature of the viscosity, effective in galactic
disks? Normal viscosity is not efficient, due to the
very low density of the gas. One could think of
macroturbulent viscosity, but the time-scales are
longer than the Hubble time at large radii, and could 
be effective only inside the central 1kpc. Instead,
if the galaxy disks develop non-axisymmetric density waves
such as spirals or bars, gravity torques are then very
effective at transfering the angular momentum outward.
 This led Lin \& Pringle (1987) to propose a 
prescription for an effective kinematic viscosity
for self-gravitating disk undergoing gravitational
instabilities.

\begin{figure}[t]
\center{\rotatebox{-90}{\resizebox{7cm}{7cm}{\includegraphics{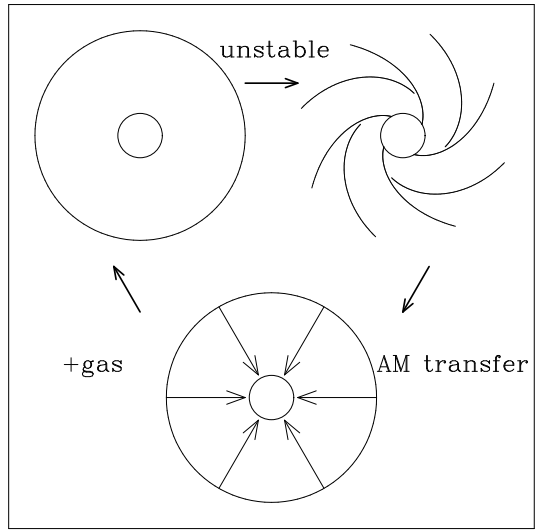}}}}
\caption{Gravitational instabilities regulate
the transfer of angular momentum, through a feedback process,
so that we can talk of "gravitational viscosity": when fresh gas has
been accreted, or radially transported inwards,
the disk becomes cold ($Q < 1$); the disk then develops waves, 
that create non-axisymmetry and gravity torques;
they transfer the angular momentum outwards
(trailing waves), and matter is driven inwards. 
The waves heat the disk, until $Q \sim 1$. 
The disk needs the presence of gas to cool down again, and
close the loop.}
\end{figure}

The basis of this prescription is described in Figure 1.
Gravitational instabilities are suppressed at small scales
through the local velocity dispersion $c$, and at large
scale by rotation. The corresponding limiting scales
are the Jeans scale for a 2D disk $\lambda_J \sim c^2/(G \mu)$,
and $\lambda_c \sim G \mu/\kappa^2$, where $\mu$ is the 
disk surface density, and $\kappa$ the epicyclic frequency.
Scales between $\lambda_J$ and $\lambda_c$ are unstable,
unless $c$ is larger than $\pi G \mu/\kappa$, or the Toomre
$Q = { {c \kappa} \over {\pi G \mu} }$ is larger than 1.
 If the disk is cold at the beginning (general case for the gas),
instabilities set in, which heat the disk until $Q \sim 1$,
and those instabilities provide the necessary angular momentum
tranfer, or viscosity, to concentrate the mass. Since
the size of the region over which angular momentum is tranferred
is $\sim \lambda_c$, and the time-scale is a rotation period,
$2 \pi / \Omega$,
the effective kinematic viscosity is $\nu \sim \lambda_c^2  \Omega$,
and the typical viscous time 
$\tau_{\nu} \sim R^2 \Omega^3 / (G^2 \mu^2)$.

Why should there be approximate agreement between the 
two time-scales, viscosity and star-formation?
This comes from the fact that the two processes depend
exactly on the same physical mechanism, i.e. gravitational
instabilities. As shown empirically by Kennicutt (1989),
the Toomre parameter $Q$ appears to control star-formation
in spiral disks. Therefore, if the regulating instabilities
have time to develop, one can expect that 
 $\tau_{\nu} \sim \tau_*$, as required for exponential
light and metallicity distribution.

In summary,
if the gravitational viscous time-scale is governing the mass
accretion, the time-scale for accretion by secular evolution
is of the order of a few dynamical times.
But there must exist a reservoir of gas in the outer parts.
 Evidence that such accretion exists can be found
in present day galaxies: the strong frequency of large-scale 
asymmetries of gas observed in the outer parts, or lopsidedness,
cannot be explained without continuous or repeated  accretions (e.g. 
Richter \& Sancisi 1994). The presence of gas warps in almost
every disk galaxies observed in HI cannot be explained
by a persistant mode, but is likely to be the manifestation
of gas accretion with a different angular momentum orientation
than the normal to the galaxy disk (Binney 1992).

\section{Galaxy Interactions}

 There is plenty of evidence that the Milky Way has
accreted mass or has even encountered merging events in the
past (e.g. Searle \& Zinn 1978, Zinn 1993). Moving groups are
detected in the halo (Majewski 1993, and this volume),
and the thick disk is best explained by an
interaction event (Bienaym\'e, this volume). This
is in line with hierachical formation.

 In hierarchical cosmological scenarios, galaxies have formed
by successive mergers of larger and larger entities. 
Primordial fluctuations of dark matter exist at all scales,
but small-scale perturbations become non-linear and collapse first, 
and are then incorporated in larger structures collapsing
later on. Small halos then merge into larger ones; the 
baryonic structures that had condensed in the halos potential
well, can also merge, with some delay. While the merging
scenario of dark halos is quite well understood and 
reproduced in simulations, or quantified analytically
by Press-Schechter formalism, the scenario concerning
the baryonic component, i.e. galaxies, is less known,
because complex physics intervenes (cooling, star formation,
feedback, IMF, metal abundance..) in supplement to
gravity. As a general statement, it is thought that
galaxy merging directly follows dark halos merging, as soon as
the virial velocities inside the merged dark structure
is below a certain threshold (roughly equal to the
escape velocity for individual galaxies, from
dynamical friction efficiency).
Since structures forming now have
quite large virial velocities, galaxy merging becomes
less and less efficient.

Galaxy interactions were undoubtedly more frequent in the
past, and many groups have tried to quantify the effect.
Already Toomre in 1977 has estimated the number of mergers
from their observed frequency at $z=0$ just taking into
account the probability of excentricities of binary orbits.
Statistics of close galaxy pairs from faint-galaxy redshift surveys
have shown that the merging rate increases as a power law with
redshift, as $(1+z)^{m}$ with $m=4\pm1.5$ (e.g. Yee \& Ellingson 1995). Lavery
et al (1996) claim that ring galaxies are also rapidly evolving,
with $m=4-5$, although statistics are still insufficient.
Many other surveys, including IRAS faint sources, 
or quasars, have also revealed a high power-law.
Governato et al (1998) from numerical simulations of standard ($\Omega = 1$)
and open ($\Omega_0 = 0.3$) CDM models find that the number
density of interacting binaries is proportional to $(1+z)^{4.2}$ and
$(1+z)^{2.5}$ respectively.

 The number of mergings for a given galaxy is still quite
uncertain observationally; the merger frequency, and
the peak merging epoch are very sensitive to the values
of universe density ($\Omega$) and the cosmological
constant ($\Lambda$, see e.g. Carlberg 1991).
The observation of relatively thin stellar
disks has been advanced as a constraining argument.

\subsection{Thickening of disks}

Galaxy interactions can easily thicken or even destroy a 
stellar disk (e.g. Gunn 1987). The fragility of stellar disks with
respect to thickening has been used by Toth \& Ostriker (1992)
to constrain the frequency of merging and 
consequently the value of the cosmological
parameter $\Omega$. They claim for instance that the Milky Way disk have
accreted less than 4\% of its mass within the last 5 10$^9$ yrs. 
Numerical simulations have tried to quantify the thickening effect
(Quinn et al 1993, Walker et al 1996). They show that the stellar disk
thickening can be large and sudden, but it is strongly moderated by 
gas hydrodynamics and star-formation processes, since the thin disk can 
be reformed continuously through gas infall. 
Velazquez \& White (1998) through N-body simulations find that
analytical derivations overstimate by factors 2-3 the disk heating
and thickening; while prograde satellites do heat the disk, 
retrograde ones produce essentially a coherent tilt of the disk.
Companion accretions therefore cause stellar
warps and asymmetric disks. 

Galaxies presently interacting
have their ratio $h/z_0$ of the radial disk scalelength $h$ to the
scaleheight $z_0$ 1.5 to 2 times lower than normal
(Bottema 1993; Reshetnikov \& Combes 1997). 
 However, since galaxies have experienced many interactions in the past,
including the presently isolated galaxies, all these perturbations, 
thickening of the planes and radial stripping, must be transient,
and disappear after an interaction time-scale, i.e. one Gyr. Present
galaxies are thought to be the result of merging of smaller units, according
to theories of bottom-up galaxy formation; a typical galaxy has accreted
most of its mass, and the existence of shells and ripples
attests of the frequency of interactions (Schweizer \& Seitzer 1992).
 This implies that the global thickness of galaxy planes can recover
their small values after galaxy interactions. Or in other words,
the disk of present day spirals has been essentially assembled at low 
redshift (Mo et al 1998).

\subsection{Merging history of the Milky Way}

 Since the number of galaxy interactions and mergers were much
larger in the past, it is likely that most accretion events 
occured long ago in the Milky Way. This is compatible
with the observation of globular clusters for example:
Unavane et al (1996) conclude that there has not been
more than 10\% of the mass accreted over the last 10 Gyr.
 The same is true for many galaxies of the Local Group,
except maybe for the SMC (Sarajedini et al 1998).

\begin{figure}[t]
\center{\rotatebox{-90}{\resizebox{8cm}{12cm}{\includegraphics{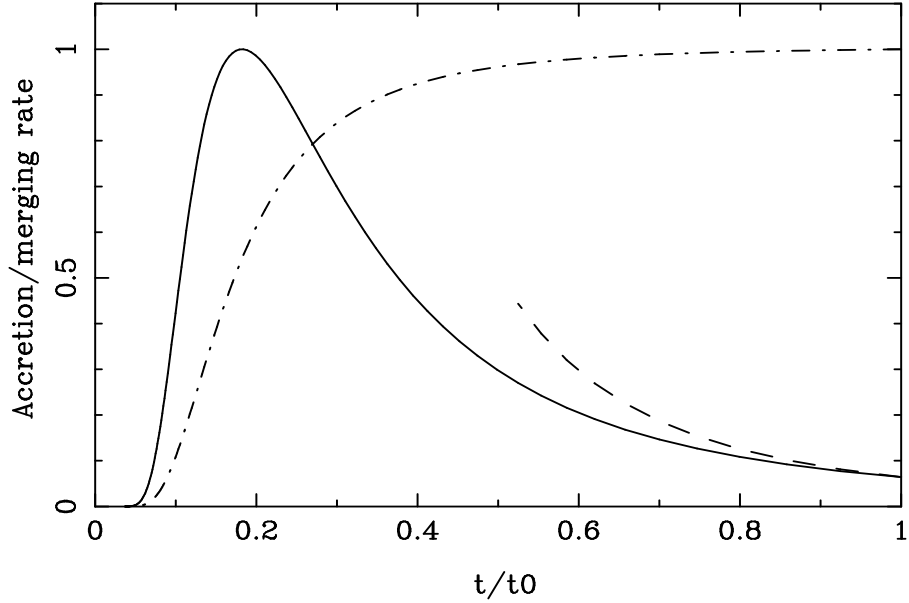}}}}
\caption{ Indicative accretion rate as a function of time, for a structure 
of the Milky Way mass (10$^{12}$ M$_\odot$ halo). 
Full line: normalised accretion rate, as expected
from the Press-Schechter formalism (standard CDM model)
for the dark halos, and modified for baryonic systems merging;
Dashed line: power law $m = 4.5$ as a 
function of (1+z) for comparison;
Dash-Dotted line: corresponding fraction of the present mass assembled
as a function of time }
\end{figure}

Although we are still 
witnessing the on-going mergers of some dwarf galaxies 
bound to the Milky Way (Sagittarius dwarf, Ibata et al 1994;
 Magellanic Clouds, Lin \& Lynden-Bell 1977, Putman et al 1998),
the bulk of the mass must have been accreted some Gyrs ago.
 Many debris stay coherent as tidal streams for the
Milky Way life-time: Johnston (1998) estimates that up
to 10\% of the sky is covered by those debris, and
this could have consequences for microlensing experiments 
(Zhao 1998). If the time-scale for mixing after stellar
accretion can be long, more than a Hubble time,
the time-scale for gas accretion and processing is
quite short (less than a Gyr). High velocity clouds
(HVCs) could represent a huge reservoir of gas around
the Milky Way, if we interprete correctly their distance.
If the latter is around 40kpc from the center, they correspond 
to a total of 10$^{11}$ M$_\odot$ of HI gas (Blitz et al 1998).
 Simulations of the clouds infalling towards the Local
Group gravity center are compatible with the observations.
HVCs will then be the analog of the Lyman-limit
absorbing clouds in absorption in front of quasars.
Embedded in dark mini-halos, they could be the building
blocks of the Galaxy (Blitz et al 1998).

 To have a statistical estimate of these accretion times,
we can use the empirical law of interactions increase 
with redshift, in $(1 + z)^m$ as described earlier. An other approach 
is to consider the probability of formation of halos of 
mass M = 10$^{12}$ M$_\odot$, such as the Milky Way.
This is obtained through the well-known Press-Schechter (1974)
formalism, or "excursion set" derivation (Bond et al 1991).
The model is based on the existence of a Gaussian random
field for initial density fluctuations, on the follow up
of their linear growth, and on spherical collapse theory.
This leads to the differential mass function:
$$
f(\sigma(M), t) = {{\delta_c}\over{\sqrt(2\pi) D(t) \sigma^3}}
exp [ -{{\delta_c^2}\over{2 D(t)^2 \sigma^2}}]
$$
where $\sigma(M)$ is the mass variance, 
$D(t)$ the linear growth factor and $\delta_c$ the critical
overdensity ${{\delta \rho}\over{\rho}}$ for collapsed structures
(extrapolated in linear theory =1.69).
The derivative of this formula with respect to time
(or redshift) gives the rate of change of density of halos
of a given mass $M$. But this is the difference between their
increase as the result of mergers of smaller halos, and their
decrease due to combining into halos of larger masses (and can
become negative).
To have the rate of merging, Carlberg (1990a) assumes that the halos 
of mass larger than $M$ come essentially from merging masses
between $M/2$ and $M$; a more exact calculation is derived
by Lacey \& Cole (1993), and leads to an analytical expression
for the merging or accretion rate as a function of time,
to form the Milky Way halo, for example. This analytical 
expression is in very good agreement with
N-body simulations (Lacey \& Cole 1994).

Once the rate of merging of halos is estimated, 
the probability of merging the baryonic systems must be taken into
account, to compare with observations,
and this introduces large uncertainties. Carlberg (1990b)
proposes a simple criterion for baryonic systems to merge:
either they merge in a dynamical time-scale if their relative 
velocity is lower than the escape speed from the systems, or 
they never merge in a Hubble time. 
 He then derives the baryonic merging rate
by considering only the fraction of the Maxwellian
distribution for velocities less than the threshold for merging.
Lacey \& Cole (1993) show
that the merging efficiency depends strongly on the 
excentricity of the initial orbits, through dynamical
friction. The results between circular orbits and highly
elongated ones (pericenter = 5\% of apocenter) can 
vary by factors 3 or more. An average estimate of the
accretion rate as a function of time
for baryonic matter is plotted in fig 2. It is very close at
low redshift to a power-law of power $m = 4.5$, the value expected
for this model universe, through numerical
simulations. In this scenario, it can be seen that most of
the Galaxy (90\% of the mass) was built 9 Gyr ago
(if $t_0$ = 15 Gyr).

\section{Conclusion}

Mass accretion can be considered to
occur through two processus: a secular acretion
from gas in the outer parts of galaxies, or in the near halo
(may be HVCs), that occurs on a dynamical time-scale; and through
galaxy interactions and mergings, that has progressively built the
galaxy in the past. Both processus can occur on widely different
time-scales, according to the environment. For instance, Low Surface
Brightness (LSB) galaxies appear unevolved systems, with large
gas fraction, low mass concentration, and their long evolution
time-scale is attributed to their poor environment (Bothun et al 1997).
 The Milky Way on the contrary, as an HSB, had a very rapid
evolution time-scale, and is likely to have accreted the bulk of its mass 
8-10 Gyrs ago. Its evolution is compatible with the statistical
mean obtained in hierachical scenarios, from the Press-Schechter 
formalism for the dark halos merging, implemented with simple
recipes for the merging of baryonic systems.


\end{document}